\documentclass[apj]{emulateapj}

\usepackage{multirow}
\usepackage{xcolor}

\shorttitle{The motion of a losing mass plasmon}           
\shortauthors{Rivera-Ortiz, Rodr\'iguez-Gonz\'alez, Hern\'andez-Mart\'inez \& Cant\'o}                          

\begin{document}    
\title{The motion of a losing mass plasmon}
\author{P.R. Rivera-Ortiz\altaffilmark{1}$^,$\altaffilmark{2}, A. Rodr\'\i guez-Gonz\'alez\altaffilmark{1}$^,$\altaffilmark{2}, L. Hern\'andez-Mart\'inez\altaffilmark{1}  \& J. Cant\'o\altaffilmark{3}}
\altaffiltext{1}{Instituto de Ciencias Nucleares, 
Universidad Nacional Aut\'onoma de M\'exico, 
Ap. 70-543, 04510 D.F., M\'exico}
\altaffiltext{2}{LUTH, Observatoire de Paris, PSL, CNRS, UMPC, Univ Paris Diderot, 5 place Jules Janssen, F-92195 Meudon, France}
\altaffiltext{3}{Instituto de Astronom\'{\i}a, Universidad
Nacional Aut\'onoma de M\'exico, Ap. 70-264, 04510 D.F., M\'exico} 
\email{pedro.rivera@correo.nucleares.unam.mx} 
\begin{abstract}
The interaction of a high velocity clump of gas has been described by 
the plasmon model, which considers balance between ram pressure and 
the internal stratified structure of the decelerated clump. In this 
paper we propose an analytical model to describe the mass loss of 
such a clump due the interaction with the environment,  describing 
its influence on the plasmon dynamics. We carry out comparisons 
between an analytic model and axisymetric gasdynamic simulations of 
plasmon evolution. From our simulations we were able to find 
the values of the friction constants $\alpha$ and $\lambda$. Comparing with the complete analytic  model from which we can infer the position and the  mass loss of the clump as function of the clump's density and the environment ratio.

\end{abstract}	
%
\section{Introduction}

The problem of a wind/molecular cloud interaction has been long studied in the past. De Young \& Axford (1967, hereafter DA) described the motion of a clump {{de}}celerated by the ram pressure and determined the lifetime of {{the}} plasmon. They applied this model to Cygnus A and concluded that analyzing the dynamics of plasmons should reduce their free parameters. It became a very popular model to explain confinement of radio lobes propagating through the
intergalactic medium (Ubachukwu, Okoye \& Onuora 1991; Daly,
1994), models of radio-loud quasars (Daly, 1995) and models of the
optical narrow-line regions of Seyfert galaxies (Taylor, Dyson \&
Axon 1992; Veilleux, Brent \& Bland-Hawthorn, 1993). Cant\'o et al., (1998; hereafter C98) rederived the plasmon solution adding the centrifugal pressure to obtain a modified plasmon profile. 

In most cases it is difficult to calculate the real age of an astronomical plasmon since there is no clear information about deceleration and most of plasmons are isolated so there is not enough information about the static medium. To solve this problem a set of several plasmons with an noticeable deceleration moving under similar restrictions is needed.

Orion BN/KL {{is an ideal laboratory to prove the plasmon solution, because}} has an almost isotropic and explosive outflow that could be produced by the non-hierarchic close dynamic interaction of a forming multiple-star system (Zapata, 2009). In this region there are more than a hundred of filamentary structures known as fingers that allow to estimate a dynamical age between 1000 years and 500 years, assuming no deceleration. Nevertheless, there is observational evidence that the longest fingers detected in H$_2$ emission are losing speed, probing their interaction with the environment (Bally et al. 2011). It is a very interesting star formation region that due { to} its distance, at 414~pc, allows us to determine its characteristics with enough detail. Therefore, we also can model the physics using theoretical and numerical models, { using some observational constrains}. Some of these models have achieved important results as determining the dynamical age and the energy of the explosive event. Nevertheless there are important questions that deserve attention and are not resolved yet, such as the real age of the event, the mechanism that can generate such distribution of the fingers, as well as their ejection velocity  since there is evidence of a drag force.

The effect of a drag force is necessary to understand the real motion of a plasmon. Several numerical simulations have shown a deceleration effect greater than the expected by ram pressure  (Yalinewich \& Sari, 2016), but it has not been deeply analyzed since cooling effects {{were not included}}. 

{  The destruction of the original clump was also considered in Raga et al. (1998) in their study of the interaction of a fast wind impinging a compact spherical cloud. They concluded that the motion is affected by the detachment of material of the cloud, which results in a limited application of their model. 

Then, the assumption that a clump has no deceleration or a deceleration according to models with constant mass, can lead to overestimate the age of astrophysical outflows. 
}

In this work, we use { the} DA solution to propose a mass loss rate for a plasmon and we obtained its equation of motion. We compare results of this analytic model with numerical simulations using Orion BN/KL plausible ejection conditions. We presented an analytic (\S~2) and numerical (\S~3) models of a deceleration of the clumps as function of ratio density when the mass loss rate is considered. We present a comparison between the analytic and numerical models and a prediction of the lifetime of clumps assuming similar conditions to the system Orion BN/KL is presented in \S~4. Finally, we present our conclusions in \S~5.

\section{Analytical model}

\subsection{De Young \& Axford's plasmon}
DA studied the problem of a clump of gas moving through an uniform environment. They found a solution (the 'plasmon' solution) based on the balance between the ram pressure of the environment and the stratified thermal pressure of the decelerating clump. For a clump of mass $M$, isothermal sound speed $c$, moving supersonically with velocity $v$ through a medium of density $\rho_a$, the plasmon adopts a pressure and density stratification given by 

\begin{eqnarray}
P=P_0 e^{-x/h}, &\quad \rho=\rho_0e^{-x/h},
\label{eqn:p-rho}
\end{eqnarray}

as a function of the position $x$ from the tip of the cloud where the pressure is $P_0$, the density is $\rho_0=P_0/c^2$, where $c$ is the isothermal speed of sound. Ram pressure with the environment with density $\rho_a$ indicates $P_0=\rho_av^2$.

In Eq. (\ref{eqn:p-rho}) 

\begin{equation}
h=\frac{c^2}{a},
\label{eqn:h}
\end{equation}

is the scaleheight of the pressure distribution, and $a$ the deceleration of the clump.

\noindent

The balance between the internal pressure and the ram pressure with the surrounding environment shapes the plasmon as, 
\begin{equation}
y=2h \arctan (e^{x/h}-1)^{1/2}.
\label{eqn:shape}
\end{equation}

Then, the mass of the moving clump of gas is related to the shape by the material enclosed by $y$,
\begin{equation}
M=\int_{0}^{\infty} \pi \rho y^2 dx=\xi_{DA} \frac{\rho_a v^2 h^3}{c^2},
\label{eqn:mass}
\end{equation}
where $\xi_{DA}=\frac{\pi}{2}(\pi^2-4)$.
\subsection{Mass loss rate}

We propose a mass loss rate per unit area, $\mu$, which depends on the density and the internal sound speed, $c$,

\begin{equation}
\mu=\lambda \rho c,
\label{eqn:mu}
\end{equation}

where { $\lambda$ is an unknown parameter  expected to be less than one.  Behind Eq. (\ref{eqn:mu}), there is the assumption that the clump looses mass at a rate per unit area proportional to the local mass density, $\rho$, and with a subsonic velocity $\lambda c$. This hypothesis have been proposed and tested, for instance, by Kahn (1980), Cant\'o and Raga (1991) and Raga, Cabrit and Cant\'o (1995) in their studies of the turbulent mixing layers produced by the interaction of interstellar outflows. An estimation of $\lambda$, in our case, is found by comparison with our numerical simulation of the problem.}

Therefore, the total mass loss rate is given by the integration of $\mu$ over the total surface of the plasmon,
\begin{equation}
\dot{M}=\int_{0}^{\infty} \mu dA=\lambda \xi_{DA} \left(\frac{8}{\pi+2}\right) \frac{\rho_a v_0^2 h^2}{c},
\label{eqn:mdot}
\end{equation}

Dividing Eq.~(\ref{eqn:mass}) by Eq. \ref{eqn:mdot} and using
Eq.~(\ref{eqn:h}) we obtain the differential equation,
\begin{equation}
\frac{1}{M} \frac{dM}{dv}=\frac{8\lambda}{(\pi+2)c}.
\label{eqn:Mv}
\end{equation}

We can use a dimensionless version of Eq.~(\ref{eqn:Mv}) with the following definitions,

\begin{eqnarray}
m=M/M_0, \quad & u= v/v_0, \quad & \alpha=\frac{8\lambda v_0}{(\pi+2)c},
\label{eqn:nodimension}
\end{eqnarray}
where $M_0$ is the initial mass and $v_0$ is the initial velocity of the plasmon
The solution to Eq.~(\ref{eqn:Mv}) is
\begin{equation}
{m}=e^{\alpha(u-1)},
\label{eqn:lnm}
\end{equation}
which relates that mass behavior as function of the plasmon speed with the constant $\alpha$.

The equation of motion of the plasmon is,

\begin{equation}
\frac{dv}{dt}=-a.
\label{eqn:motion}
\end{equation}

Solving Eq. (\ref{eqn:mass}) for $h$, Eq. (\ref{eqn:h}) for $a$, and substituing in Eq. (\ref{eqn:motion}), we find the equation of motion in a non-dimensional form

\begin{equation}
\frac{du}{d\tau}=-\left(\frac{u^2}{m}\right)^{1/3},
\label{eqn:du/dtau}
\end{equation}

where

\begin{eqnarray}
\tau=t/t_0, \quad & t_0=\left(\frac{M_0v_0}{\xi_{DA}\rho_a c^4}\right)^{1/3}.
\label{eqn:time}
\end{eqnarray}

Eq. (\ref{eqn:du/dtau}), {  together with Eq. (\ref{eqn:lnm}),} has the formal solution,
\begin{equation}
\tau=\int_{u}^{1} u^{-2/3}e^{-\frac{\alpha}{3}(1-u)} du.
\label{eqn:fulltau}
\end{equation}


The position of the clump after ejection $R$ is found by solving the kinematic equation,

\begin{equation}
\frac{dR}{dt}=v
\label{eqn:R}
\end{equation}

Defining $r=R/(v_0t_0)$ and combining it with Eq. (\ref{eqn:R}), we find the solution
\begin{equation}
r=\int_u^1 u^{1/3}e^{-\frac{\alpha}{3}(1-u)}du.
\label{eqn:r}
\end{equation}

Then Eq. (\ref{eqn:lnm}), Eq. (\ref{eqn:fulltau}), and Eq. (\ref{eqn:r}) give the mass $m$, velocity $u$ and position $r$ of the clump after a time $\tau$ of ejection, using $u$ as the free variable in the interval [0,1].

The clump halts at a finite time $\tau_f$ and finite distance $r_f$ with finite mass $m_f$. These limits are determined by the condition $u=0$ in Eq. (\ref{eqn:lnm}), Eq. (\ref{eqn:fulltau}) and Eq. (\ref{eqn:r}) and are functions of $\alpha$ only.
We can find useful approximations in the limits $\alpha \ll 1$:
\begin{eqnarray}
\tau_f \simeq 3\left(1-\frac{\alpha}{4}\right),\\
r_f \simeq \frac{3}{4}\left(1-\frac{\alpha}{7}\right),
\label{eqn:alfa1}
\end{eqnarray}
{ which} are consistent with the DA solution. {  Furthermore, for $\alpha=0$ in Eq. (\ref{eqn:fulltau}) and Eq. (\ref{eqn:r}) we recover the solution of DA.}

Also, there are some interesting results concerning the stopping time $\tau_f$ and distance $r_f$ that deserve to be highlighted.

First, we must notice that independent of the physical characteristics of the original clump (shape, density structure or internal sound speed), the initial interaction with the medium through which it moves will modify these characteristics to those of a plasmon. That is, its shape will be transformed to that given by Eq. (\ref{eqn:shape}), its pressure and density stratification given by Eq. (\ref{eqn:p-rho}) and so on. This transformation is actually accomplished by a reverse shock that moves inside the original clump, changing it into a plasmon.

As we have seen above the structure of the plasmon is highly dependent on its internal sound speed i.e., on its temperature). Let us assume that the temperature of the newly formed plasmon is the one left by the reverse shock that moved through it. For simplicity, let also assume that this shock is planar and strong. In the Appendix we show that the corresponding isothermal sound speed is,
\begin{equation}
c=v_0\left( \frac{\gamma-1}{2}\right)^{1/2}{\beta},
\label{eqn:c}
\end{equation}
where $\beta=\sqrt{\rho_a/\rho_{cl}}$ is the square root of the ratio of the density of the environment and the density of the original clump.
Using Eq.  (\ref{eqn:c}) in Eq. (\ref{eqn:nodimension}) we find
\begin{equation}
\label{eq:lambda}
\alpha=\frac{8\lambda}{\pi+2}\sqrt{\frac{2}{\gamma-1}}\left(\frac{1}{\beta}\right),
\end{equation}
which is independent of the velocity $v_0$ and depends only on the ratio $\beta$.
Next, let us consider the time $t_f$ for the clump to stop. It is given by 
\begin{equation}
    t_f=t_0 \tau_f(\alpha),
\end{equation}
where $t_0$ is defined by Eq. (12) as,
\begin{equation}
    t_0=\left( \frac{M_0 v_0}{\xi_{DA} \rho_a c^4} \right)^{1/3},
\end{equation}
or
\begin{equation}
    t_0=\left[ \frac{M_0}{\xi_{DA} \rho_a } \left( \frac{v_0}{c} \right)^4 \right]^{1/3} \frac{1}{v_0}, 
    \label{eqn:stop}
\end{equation}
{ and corresponds to the timescale used by DA in their solution to estimate the lifetime of a plasmon (Eq. (\ref{eqn:uDA}))}. 

As shown in the Appendix (see also Eq. (20)) the ratio $v_0/c$ is only function of the contrast density $\beta$ and independent of the velocity $v_0$. Thus, given the ratio $\beta$, the time $t_0$ and therefore the time $t_f$ for the clump to stop diminish as the initial velocity of the clump increases. This is an unexpected result: the time for stopping a stripping clump is inverse with its initial velocity. Faster clumps stop earlier independent of their size. 
Now, let us consider the stopping distance $R_f$. This is given by,
\begin{equation}
    R_f=v_0t_0r_f(\alpha),
\end{equation}
From the discussion above, the product $v_0t_0$ results independent of $v_0$, and thus $R_f$. 
Then, clumps with the same ratio {  $\beta$} stop at the same distance from the injection point, independent of either its initial velocity or size.

\section{Axisymmetric simulations of plasmon evolution}     

\subsection{The numerical setup}
In order to validate the analytical model, we have computed axisymmetric numerical simulations with the full radiative gasdynamic equations.
We used the {\sc Walkimya 2D} code (see Castellanos-Ram\'irez et al. 2018 and Esquivel et al. (2010)) to perform all numerical simulations. The code solves the hydrodynamic equations and chemical networks on a two dimensional Cartesian adaptive mesh, using a second order finite volume method with HLLC fluxes (\citealt{Toro1994}).

The adaptative mesh consist of four root blocks of $16 \times 16$ cells, with 7 levels of refinement, yielding a maximum resolution of $4096 \times 1024$ (axial $\times$ radial) cells. { {The boundary conditions used on the symmetry axis are reflective and the other ones are outflows.}} The size of the mesh is large enough so that the choice of outer boundaries does not affect the simulation. 

The energy equation includes the cooling function described by Raga \& Reipurth (2004) for atomic gas and for lower temperatures we have included the parametric { molecular cooling function} presented in Kosi\'nsky \& Hanasz, 2007,
\begin{equation}
\label{eq:coolingmol}
\Lambda_{\rm mol}(T)=L_1 \cdot T^{\epsilon_1}+L_2 \cdot \exp \left ( -\frac{c_*}{(T-T_*)^{\epsilon_2}}\right) \;\;,
\end{equation}
for $T < 5280$~K, where, $\epsilon_1=10.73$, $\epsilon_2=0.1$ , $L_1=4.4 \times 10^{-67}$~ erg~ cm$^3$~ s$^{-1}$~K$^{-\epsilon_1}$, $L_2=4.89 \times 10^{-25}$erg~ cm$^3$~ s$^{-1}$, $c_*=3.18$ K$^{\epsilon_2}$ and 
$T_*=1.$~K. The total radiative energy for temperatures lower than 5280~K is given by,
\begin{equation}
\label{eq:lradmol}
L_{rad,mol}=n_{\rm gas} \cdot n_{\rm CO}*\Lambda_{\rm mol}(T)
\end{equation}
where, $n_{\rm gas}$ and $n_{\rm CO}$ is the numerical density of the gas and the CO molecule, respectively.

We have also considered the heating of the gas via cosmic rays, using the heating rate presented in Henney et al. (2009), 
\begin{equation}
\label{eq:heat}
\Gamma_{\rm crp}=5\times 10^{-28}n_{\rm H} {\rm ~ erg~s^{-1}}
\end{equation}
where, $n_{\rm H}$ is the numerical density of the all the hydrogen species.

\begin{figure*}[!ht]
\centering
\includegraphics[width=1.5\columnwidth]{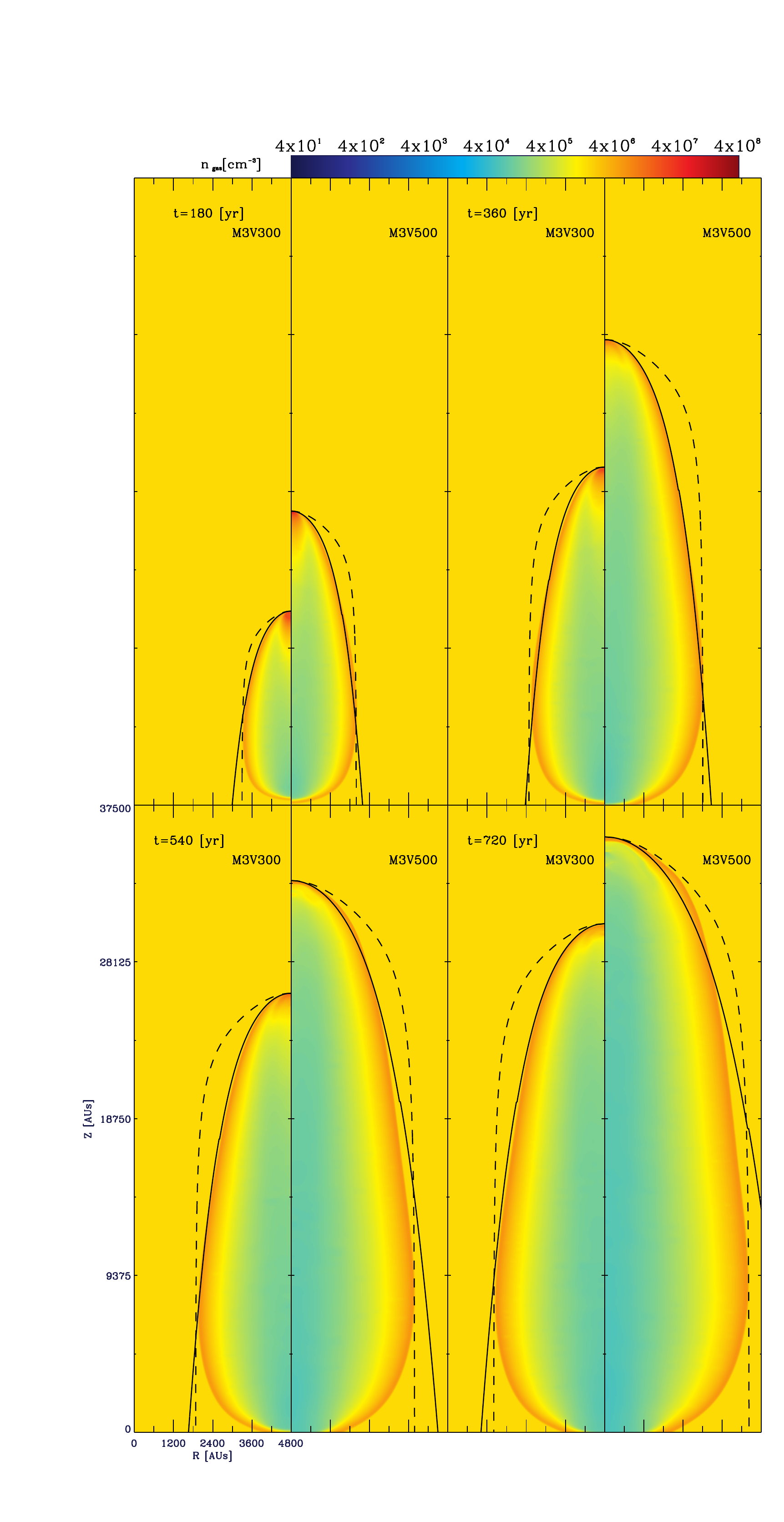}
\caption{Snapshots from the numerical simulations showing the numerical density. 
Each panel compares two models with initial velocity 
of 300 (left) and 500 (right) km~s$^{-1}$ at $t=$ 
$180,$ $360,$ $540$ and $720$ years. Lines are the 
analytical fit, {{dashed lines of DA, solid lines our fits.}}}
\label{fig:rhomaps}
\end{figure*}
\subsection{Numerical models of the plasmon evolution}

In order to study the deceleration of a high velocity clump we use compatible parameters with the ejection of Orion Fingers in Orion BN/KL.  We have { run} numerical simulation assuming that the computational domain was initially filled by a homogeneous, stationary ambient medium with temperature $\rm T_{env}=$ 100 K {  and various densities (see below)}. The numerical integration had a domain 
with physical size of 48000$\times$12000~AU on each side, with a maximum resolution (along the two axes) of $11.7$~AU.
We carried out time integration from t$_i$=0 to t$_f$=1000~yr, and the clump is released at z=700~AU for all models. An estimation of the initial mass in each of the clumps is $\rm m_{cl}=0.03$~M$_\odot$ since the total mass of the moving gas in the $\sim$400 fingers in Orion~BN/KL is about 8~M$_{\odot}$ (Bally, 2016). { Also, the observed transverse size of the fingers is about 400 AU.}

In the numerical models,  the initial clump is imposed {  as} a sphere of radius $\rm R_{cl}$=50~AU, corresponding to 4~pixels at the maximum resolution of the adaptive grid and with an uniform density of $\rm n_{cl}$=1$\times$ 10$^{10}$ cm$^{-3}$. 
{ Since the initial clump is out of equilibrium, it increases its size to about 400 AU from the first output, and then the density structure of a plasmon arises.}

We have computed ten simulations of the clumps, varying the density 
of the interstellar medium and the velocity {{at}} which the clump was thrown (see Table~\ref{tab:sim}).

\begin{table}[!ht]
\begin{center}
\caption{Initial conditions of the numerical models}
\label{tab:sim}
 \begin{tabular}{ccc}
\hline \hline
\multicolumn{1}{c}{Models} &
\multicolumn{1}{c}{Environment} &
\multicolumn{1}{c}{clump}\\
\hline
\multicolumn{1}{c}{} &
\multicolumn{1}{c}{n$_{\rm a}$} &
\multicolumn{1}{c}{$v_0$} \\ 
  
\multicolumn{1}{c}{} &
\multicolumn{1}{c}{[cm$^{-3}$]} &
\multicolumn{1}{c}{[km~s$^{-1}$]} \\
\hline
M1V300/M1V500 & 1.0$\times$10$^6$    & 300/500 \\
M2V300/M2V500 & 3.16$\times$10$^6$   & 300/500 \\
M3V300/M3V500 & 1.0$\times$10$^7$    & 300/500 \\
M4V300/M4V500 & 3.16$\times$10$^7$   & 300/500 \\
M5V300/M5V500 & 1.0$\times$10$^8$    & 300/500 \\
\hline
\end{tabular}

\end{center}
\end{table}

One of the main hypothesis of our analytic model is that the early interaction of the original clump with the environment will modify its initial characteristics (shape, density stratification or sound speed) to those of a plasmon. { The sound speed of the moving clump is calculated using the internal temperature, which is about 15~K and is in the order of magnitude of the sound speed obtained with Equation \ref{eqn:cmodel} }

In order to illustrate the numerical simulation results, in Figure~(\ref{fig:rhomaps}) we
present the density maps for models M3V300 and M3V500 (left and right panels, respectively) at 
evolutionary time of 180, 360, 540 and 720 yr, top left, top right, bottom-left and bottom right panels, 
respectively. The solid lines, in all the panels, are the analytical fit of the plasmon shape, {{Eq.~(16) and Eq.~(17)}} 
presented in C98 and the dashed lines are also the plasmon shape's fit obtained by DA in their {{Eq.~(2).}} 
In both models, the plasmon shape expected by the DA equation is wider than the shape of the plasmon's head 
obtained in the numerical simulations. For model M3V500 (the right panels of Figure~(\ref{fig:rhomaps})) the 
plasmon shape proposed by C98 is in very good agreement with the numerical simulations, at 
least {  up to} t$\sim$500 yr. After this time, the plasmon (of the model M5V300) is rapidly decelerated and { the 
bow shock changes in a different shape 
that the {  one} proposed by C98}. The models with
lower initial velocity, model M3V300, does not have {  an} appreciable {  deceleration}  and the 
numerical simulation shape is {  in agreement} with the C98 prediction. 

\begin{figure}[!ht]
\includegraphics[width=0.99\columnwidth]{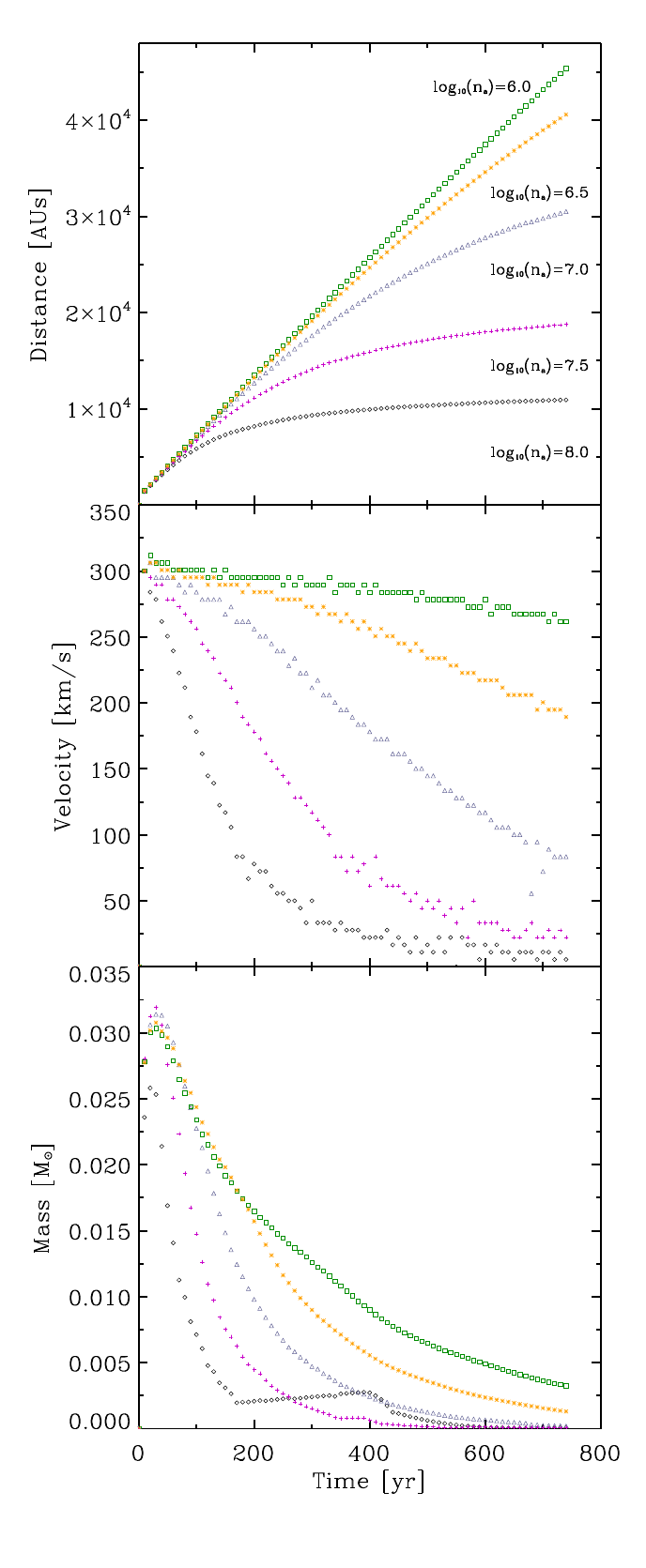}
\caption{The top, middle and bottom panels shows the position, velocity and mass
as faction of time by the numerical model with initial velocities of 300 km~s$^{-1}$, respectively. In 
each of this panels we plot the results obtained for the models evolving with logarithmic interstellar 
medium densities of 6, 6.5, 7, 7.5 and 8 { using 
green squares, yellow asterisks, blue triangles, magenta crosses and black diamonds symbols, respectively.} }
\label{fig:v300}
\end{figure}

\begin{figure}
\includegraphics[width=\columnwidth]{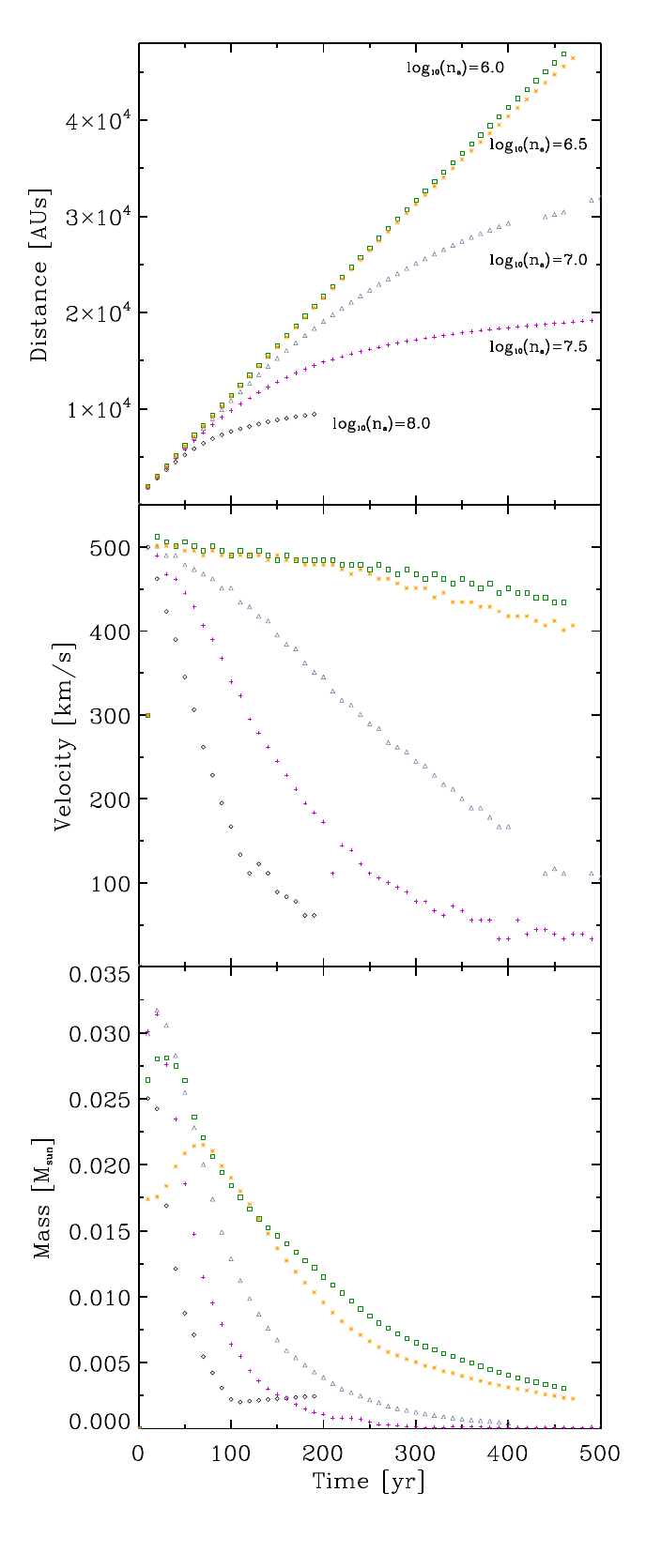}
\caption{The same as Figure~\ref{fig:v300} but for models with initial velocity of 500~km~s$^{-1}$.}
\label{fig:v500}
\end{figure}

{ Other prediction of our model is that the { dimensionless} mass $m$ of the clump is related to its { dimensionless} velocity $u$ by Eq. (\ref{eqn:lnm}). We can test this prediction.} The top panel of Figure~\ref{fig:v300} shows the position as function of time by plasmon for the models 
with initial velocity of 300~km~s$^{-1}$ (see Table~(\ref{tab:sim})). The squares, asterisk, triangles, plus and 
diamond symbols are drawing the results obtained for the models evolving with logarithmic interstellar medium 
densities of 6, 6.5, 7, 7.5 and 8, respectively. As we can see, the position is {  smaller} for models that 
{  evolve} in denser environments, it means 
the deceleration or decrease on the plasmon's velocity, as function of time  (see~middle panel of Figure~(\ref{fig:v300})) 
is larger in models with larger ram pressure (Eq.~(\ref{eq:ram})). And in the bottom panel of this figure, we presented the 
mass of the clump as function of time. We calculated the mass considered the gas into the sphere of 50~AU 
of radii since the clump position and we also note that the denser interstellar medium produce a larger mass 
loss rate in the clumps moving in {{environments}} with uniform density and temperatures.
In the same way, Figure~(\ref{fig:v500}) shows the position, velocity and mass as function of 
time of the numerical simulations {  for} a larger initial velocity, 300~km~s$^{-1}$. 
The {  results for} the distance, velocity and mass are very similar to those found in the models with 
initial velocity of {  300} km~s$^{-1}$. However, the deceleration for the models with initial velocity 
of 500 km~s$^{-1}$ is larger than {  those for} the model{  s} with $v_0$=300~km~s$^{-1}$, 
and lifetime of the faster clumps is smaller than {  those for} the lower ones, {{as we predicted in our Eq.~(\ref{eqn:stop}).}}

\begin{figure}

\includegraphics[width=\columnwidth]{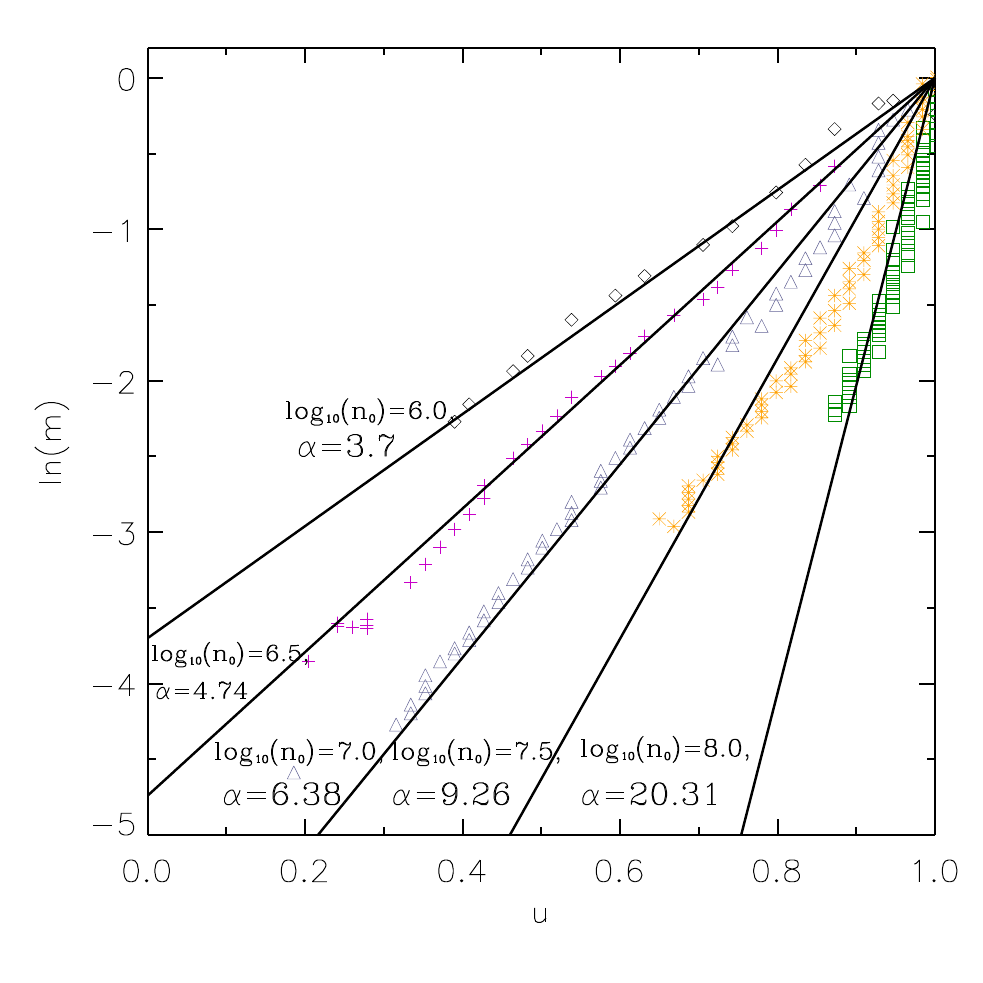}
\caption{Mass of the clump as function of the clump's velocity (dimensionless). The nomenclature of the symbols is the same as in the Figure~(\ref{fig:v300}) and 
in solid lines we plot the fit for each of the models with initial velocities of 300~km~s$^{-1}$.}
\label{fig:mvsu300}
\end{figure}

\begin{figure}
\includegraphics[width=\columnwidth]{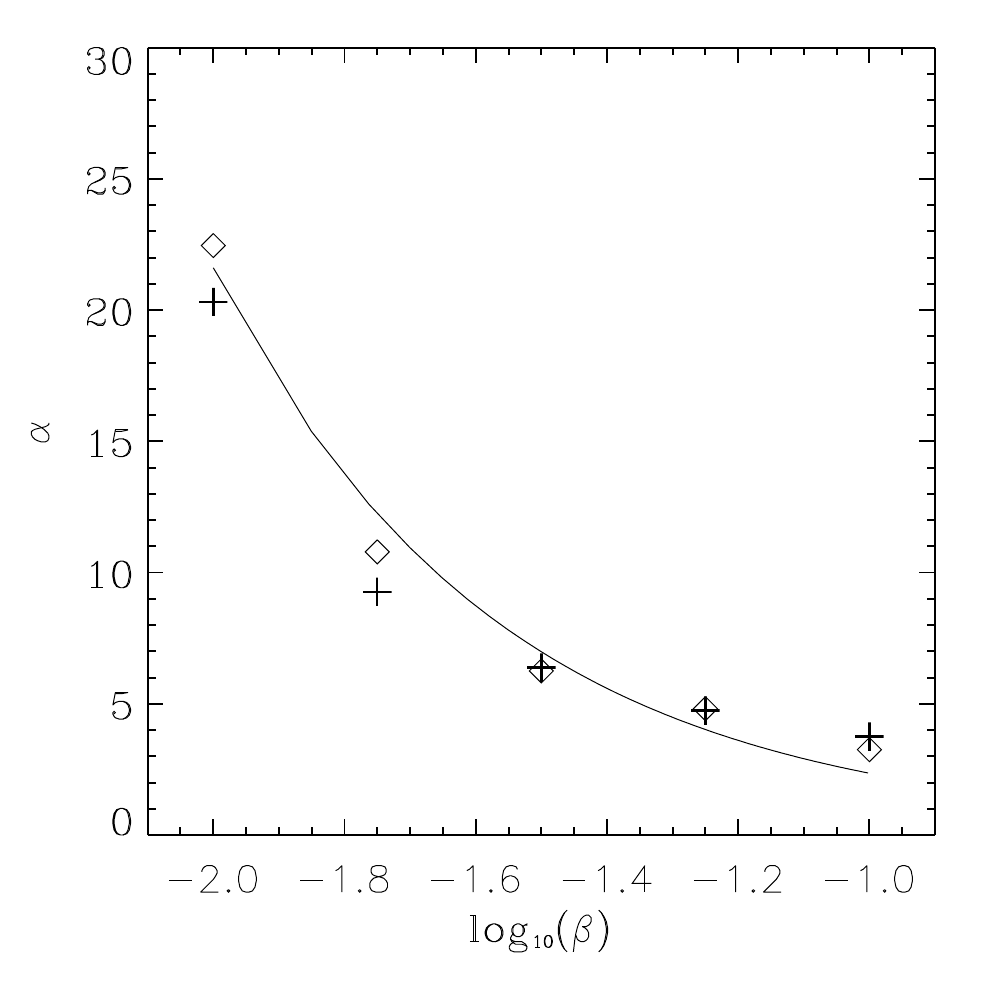}
\caption{The constant $\alpha$ as function of contrast density ($\beta$). The plus 
and diamond symbols are the $\alpha$ values for models with $v_0$=300 
and 500~km~s$^{-1}$, respectively and the solid line is the best fit of $\lambda$ (see
Eq.~\ref{eq:lambda}), $\lambda$=0.0615.}
\label{fig:alfas}
\end{figure}

Using the dimensionless mass of the clump and velocity from our numerical simulation in the Eq.~(\ref{eqn:fulltau}) we fitted the $\alpha$ value for all the numerical model. Figure~(\ref{fig:mvsu300}) shows the logarithm of the 
mass of the clump as function of velocity (dimensionless), for all the models with initial velocity 
of 300~km~s$^{-1}$, we use the same nomenclature for the symbols as in the Figure~(\ref{fig:v300}), and the solid lines
are the fit for the models, M1V300, M2V300, M3V300, M4V300, M5V300 and M6V300.\\
The $\alpha$ values for all the models presented here are plotted in
 Figure~(\ref{fig:alfas}). The plus and diamond symbols are the $\alpha$ values 
 fitted for modes with $v_0$=300 and 500~km~s$^{-1}$, respectively. In order to 
 obtain the value for the constant $\lambda$ (see Eq.~(\ref{eq:lambda})), we
have fitted the $\alpha$ values as function of the contrast density $\beta$ {  to} our numerical simulation (solid line in this figure). {  The best fit gives $\lambda= 0.0615$}.
Notice that the $\alpha$ values are only function of contrast density and these values are not dependent of the initial velocity or other parameters of the cloud, as described by Eq.~(\ref{eq:lambda}). $\lambda$ is a constant that is independent of the physical properties of the interstellar medium or 
clump gas.
\section{Prediction of evolutionary physical properties of the plasmon}

{  The solution for a constant mass plasmon can be found from Eq. (\ref{eqn:fulltau}) with $\alpha=0$. The results are, 

\begin{equation}
u=\left(1-\frac{\tau}{3}\right)^3,
\label{eqn:uDA}
\end{equation}

for the velocity, and

\begin{equation}
r=\frac{3}{4}\left[ 1- \left( 1-\frac{\tau}{3} \right)^4  \right]
\label{eqn:rDA}
\end{equation}

for the position.}

Therefore, in the approximation of DA the lifetime of a plasmon is $t_f=3\cdot t_0$  (see Eq.~(\ref{eqn:uDA})). 
When {  the} mass loss rate is taking into account the plasmon's motion is changed, and the Eq.~(\ref{eqn:fulltau}) can be integrated numerically to obtain $u$ and the dimensionless position $r=x/x_0$ with $x_0=v_0 t_0$. {{It is important to recall that C98 included the centrifugal pressure, which can affect the plasmon shape. This effect {  was} not included in DA {  neither by us}.}}

\begin{figure}[!ht]
\includegraphics[width=\columnwidth]{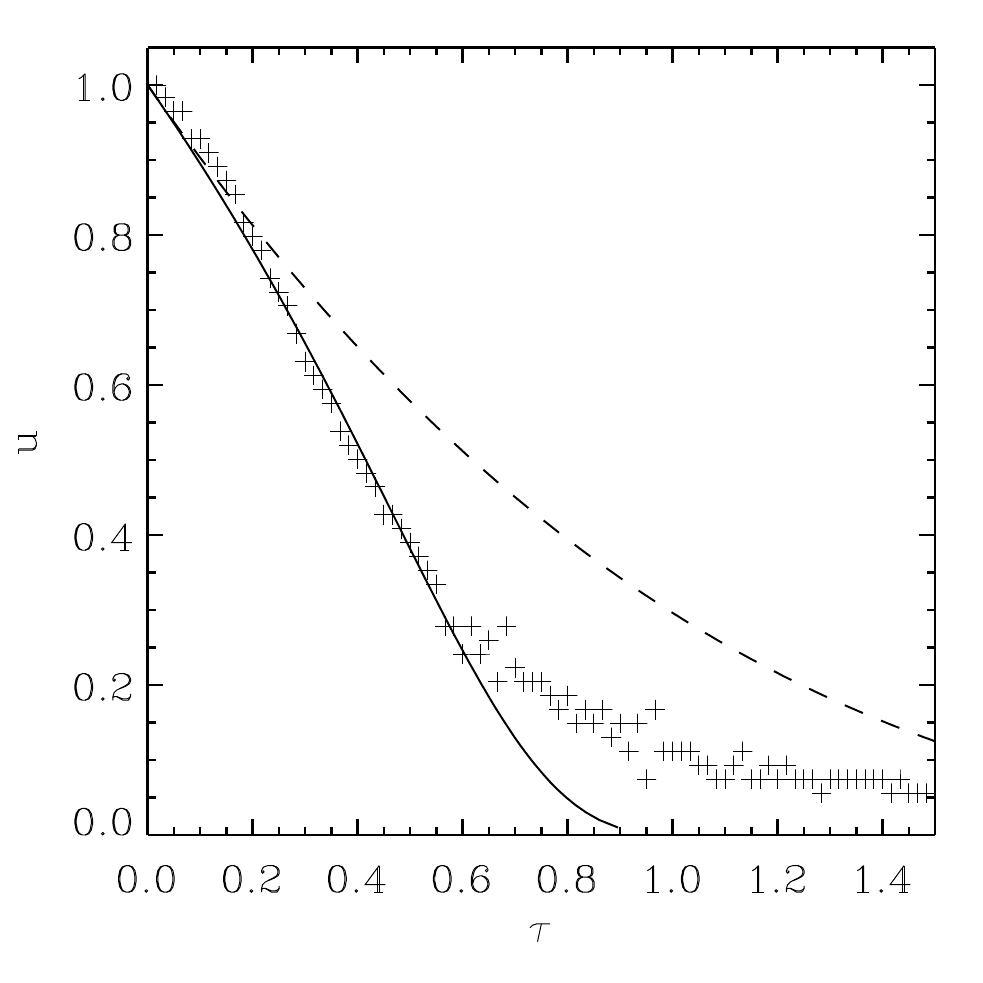}
\caption{Dimensionless velocity, $u$, vs dimensionless time, $\tau$, for the model with $\log \beta=-1.25$. Solid line represents the solution to Eq. \ref{eqn:fulltau}, dashed line is the De Young and Axford prediction in Eq. \ref{eqn:uDA} and crosses are the numerical simulation data normalized with $v_0=300$~km s$^{-1}$ and $t_0=600$~yr}
\label{fig:ut}
\end{figure}

{{Finally}}, we use our numerical simulations to probe our models and their limitations. Each simulation has physical units, so they have to be normalized {  with} 
$v_0$, $t_0$ and $x_0$. $v_0$ is obtained directly from the initial 
conditions, $t_0$ is obtained from a fit of the velocity data and $x_0$ 
comes from a similar fit {{of the }}position data. \\
Figure~(\ref{fig:ut}) shows the dimensionless velocity as function of dimensionless time.  From this figure, one can
see that the DA solution {  agrees with the numerical results of the
model M4V300 only for $\tau \leq $0.2 (crosses)}. However, the semi-analytical solution, solid line {  is in agreement with the numerical model up to $ \tau=0.6$}. Notice, that after $\tau=0.6$ the values of $u$, for the numerical simulation, tends to {  a} constant. There are numerical uncertainties that 
lead to overestimate the velocity, since, as the plasmon losses mass, it is difficult to determine its position and therefore {  its} velocity.

\begin{figure}[!ht]
\includegraphics[width=\columnwidth]{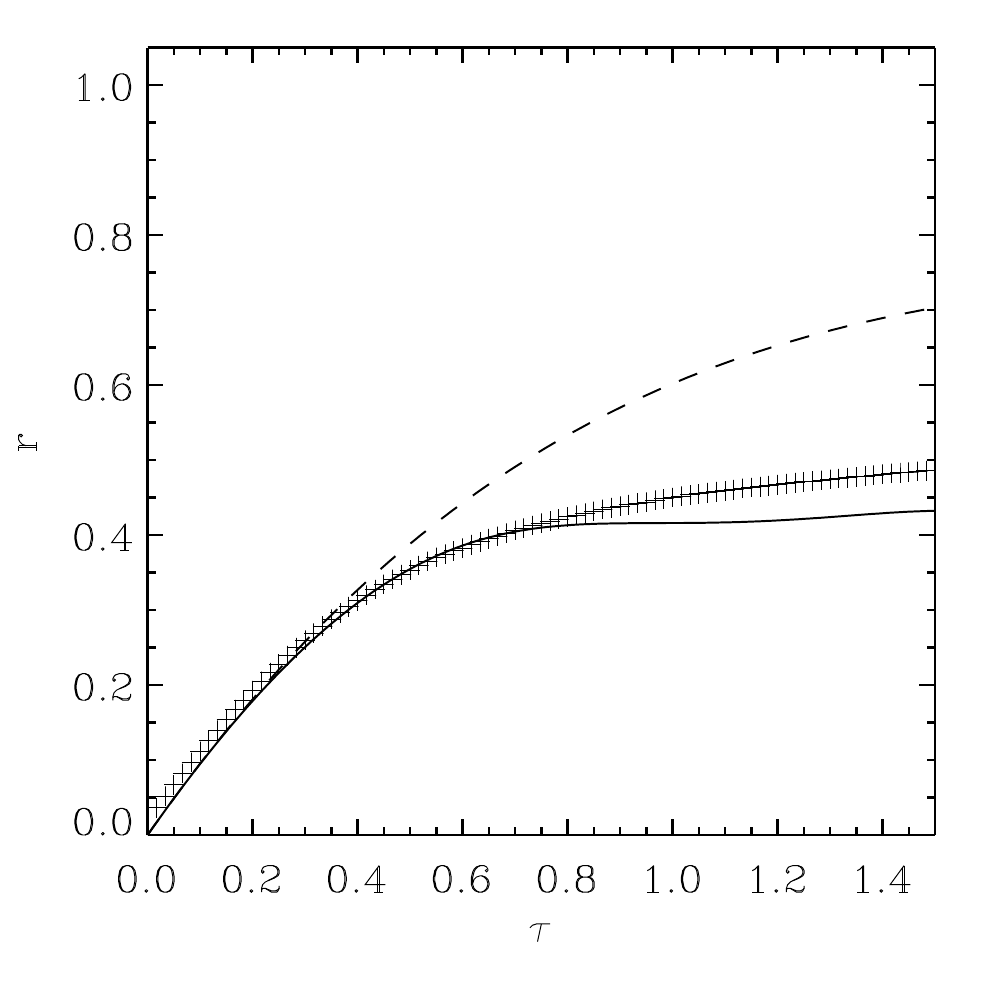}
\caption{Dimensionless position $r$ vs dimensionless time $\tau$ for the model with $\log \beta=-1.25$. Solid line represents the solution to Eq.~(\ref{eqn:fulltau}), dashed line is the De Young and Axford prediction in Eq.~(\ref{eqn:rDA}) and crosses are the numerical simulation data normalized with $x_0=38000 $AU and $t_0=600$yr}
\label{fig:rt}
\end{figure}

Figure~(\ref{fig:rt}) shows the dimensionless position as function of dimensionless time. {  The } DA solution, semi-analytic solution and numerical data are represented {  in this figure. The} DA solution is similar to {  the} numerical data {  for $\tau \leq 0.4$, while for the semi-analytical solution the agreement extends up to $\tau \simeq 0.6$ }. In the case of semi-analytical solution this time is as long as $\tau=0.8$, {  similar stop distance while the DA model predicts a larger distance}

Finally,  the analytic $t_0$ and $x_0=v_0t_0$ obtained from Eq. (\ref{eqn:time}) and the numerical $t_0$ and $x_0$, for all the models presented in {  Table~(\ref{tb:compare300}) and Table~(\ref{tb:compare500}) for initial velocities 300~km~s$^{-1}$  and 500~km~s$^{-1}$, respectively}.

\begin{table}[!ht]
\centering
\caption{Analytic and numerical scale length $x_0$
and time $t_0$ for models with $v_0=300$ km s$^{-1}$}
\label{tb:compare300}
\begin{tabular}{c c c c c}
\hline \hline
\multirow{3}{*}{$\log \left(\frac{n_{\rm a}}{{\rm [cm^3]}}\right)$} & \multicolumn{2}{c}{Analytical}         & \multicolumn{2}{c}{Numerical} \\ \cline{2-5} 
                                  & \multicolumn{1}{c}{$x_0$}     & $t_0$ & $x_0$           & $t_0$        \\
                                  & \multicolumn{1}{c}{[AU]}      & [yr]    & [AU]     & [yr]           \\ \hline
6                                 & \multicolumn{1}{c}{$1.14\times 10 ^6$} & 18369 & $1.2\times 10^ 6$        & 18400        \\  
6.5                               & \multicolumn{1}{c}{363868}    & 5868  & 200000          & 3000         \\  
7.                                & \multicolumn{1}{c}{117000}    & 1900  & 90000           & 1400         \\  
7.5                               & \multicolumn{1}{c}{38231}     & 616   & 38000           & 600          \\  
8                                 & \multicolumn{1}{c}{12762}     & 205   & 18000           & 250          \\  
\hline
\end{tabular}
\end{table}

{{From the analytical solution, we can see}} that the final position
(the scale length) are only function of the contrast density {{and}} it is not related with the velocity {{at which}} the clump was thrown, {{see Table~(\ref{tb:compare300}) and Table (\ref{tb:compare500}}}). However, {{the lifetime of the}} plasmon or clump is {  depends on} the initial velocity and the density contrast. 
The plasmon that were faster initially suffer a higher deceleration.

\begin{table}[!ht]
\centering
\caption{Analytic and numerical scale length $x_0$ \\
and time $t_0$ for models with $v_0=500$ km s$^{-1}$}
\label{tb:compare500}
\begin{tabular}{ccccc}
 \hline \hline
\multirow{3}{*}{$\log \left(\frac{n_{\rm a}}{{\rm [cm^3]}}\right)$} & \multicolumn{2}{c}{Analytical}         & \multicolumn{2}{c}{Numerical} \\ 
\cline{2-5} 
                                  & \multicolumn{1}{c}{$x_0$}     & $t_0$ & $x_0$           & $t_0$        \\
                                  & \multicolumn{1}{c}{[AU]}      & [yr]    & [AU]     & [yr]           \\ \hline
6                                 & \multicolumn{1}{c}{$1.14\times 10 ^6$} & 11022 & $1.1\times 10^ 6$        & 10000        \\ 
6.5                               & \multicolumn{1}{c}{363868}    & 3500  & 300000          & 2700         \\ 
7.                                & \multicolumn{1}{c}{117000}    & 1134  & 90000           & 850         \\ 
7.5                               & \multicolumn{1}{c}{38231}     & 370   & 40000           & 400          \\ 
8                                 & \multicolumn{1}{c}{12762}     & 123   & 19000           & 165          \\ \hline
\end{tabular}
\end{table}

\section{Conclusions}

We have used the plasmon solution obtained by {{DA}}  and the solution presented in C98 to propose an analytical solution of the plasmon's deceleration when mass is considered.

This leads to interpret mass as a function of the plasmon velocity related by a constant $\alpha$.  This $\alpha$ can be interpreted as a friction {  coefficient. We calculate} its dependence  on the density contrast between the plasmon and the surrounding environment.

Several numerical simulations were performed trying to compare the validity {  of our analytic model. An estimation of $\lambda=0.0615$ was found}.

The lifetime obtained from the simple plasmon model is greater than the expected by our losing mass considerations. The deceleration obtained by this method is more likely to be responsible for the age discrepancy in astronomical flows as the Orion fingers. Also, it is important to notice that a plasmon with greater ejection speed has a shorter lifetime, which can be observed on simulations.

The final length of a plasmon is not related to its shape and depends on the initial conditions of the plasmon.

\begin{acknowledgements} 
We acknowledge support from PAPIIT-UNAM grants IN-109518 and IG-100218. P.R.R.-O. acknowledges
scholarship from CONACyT-M\'exico and financial support from COZCyT. We thank an anonymous referee for helpful comments and corrections.
\end{acknowledgements}

\appendix

\section{Speed of sound}
\begin{figure}[!ht]
\includegraphics[width=0.8\columnwidth]{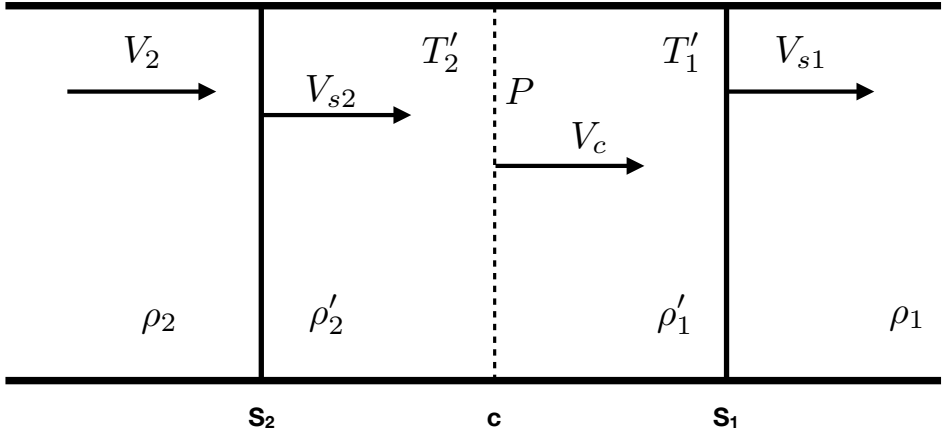}
\caption{Scheme of the flow configuration produced by the interaction of a highly supersonic flow 2 with a gas at rest 1.}
\label{fig:shock}
\end{figure}

Consider a supersonic flow with velocity $v_2$ and density $\rho_2$ interacting with a medium at rest with density $\rho_1$. The interaction produces two shocks $S_1$ and $S_2$ (see Figure (\ref{fig:shock})). Between the shocks there is a growing region that has an uniform velocity $v_c$ and uniform pressure P. $S_1$, the forward shock, moves with velocity $v_{S1}$ and runs into the medium at rest, accelerating it to the velocity $v_c$, while $S_2$, the reverse shock, moves with velocity $v_{S2}$ into the impinging flow decelerating it to the same velocity $v_c$. The region has two parts: one has density $\rho'_2$ and temperature $T'_2$ and is filled by shocked flow 2, while the other part is filled by shocked medium 1 and has density $\rho'_1$ and temperature $T'_1$. Note that the pressure of both regions is however the same. These two regions are separated by a contact discontinuity $C$. We further assume that the shocks are strong and parallel. On a frame of reference moving with shock $S_2$, we can write,

\begin{equation}
\rho'_2=\frac{\gamma+1}{\gamma-1}\rho_2,
\end{equation}
\begin{equation}
v^{'}_{2}=\frac{\gamma-1}{\gamma+1}(v_2-v_{S2})=v_c-v_{S2},
\end{equation}
and
\begin{equation}
\label{eq:ram}
P=\frac{2}{\gamma+1}\rho_2(v_2-v_{S2})^2,
\end{equation}


where $v'_2$ is the post-$S_2$ shock flow velocity in this frame of reference and $\gamma$ is the ratio of specific heats.

Now, in a frame of reference that moves with shock $S_1$, the jump conditions across the shock gives,


\begin{equation}
\rho'_1=\frac{\gamma+1}{\gamma-1}\rho_1,
\end{equation}
\begin{equation}
v_{1}^{'}=\frac{\gamma-1}{\gamma+1}(-v_{S1})=v_c-v_{S1},
\end{equation}
and
\begin{equation}
\label{eq:ram}
P=\frac{2}{\gamma+1}\rho_1(-v_{S1})^2,
\end{equation}


where $v'_1$ is the post-$S_1$ shock velocity in this frame of reference.

From (A3) and (A6) we find

\begin{equation}
v_2-v_{S2}=\beta \, v_{S1},
\end{equation}

where $\beta=(\rho_1/\rho_2)^{\frac{1}{2}}$.

Combining (A7) with (A2), (A5) and (A6) we find


\begin{equation}
v_c=\frac{v_2}{1+\beta}
\end{equation}

\begin{equation}
v_{S1}=\frac{\gamma+1}{2(1+\beta)}v_2
\end{equation}

\begin{equation}
v_{S2}=\frac{2+\beta(1-\gamma)}{2(1+\beta)}v_2
\end{equation}

\begin{equation}
P=\frac{\gamma+1}{2(1+\beta)^2}\rho_1 v_2^2
\end{equation}


Finally, the isothermal sound speed behind shock $S_2$ is, 

\begin{equation}
c_2=\sqrt{\frac{P}{\rho'_2}}
\end{equation}

\begin{equation}
c_2=\left(\frac{\gamma-1}{2}\right)^{1/2}\left(\frac{\beta}{1+\beta}\right)v_2
\label{eqn:c2}
\end{equation}


where we have used (A1) and the definition of $\beta$.

We can use Eq. (A12) to estimate the sound speed of the gas that was left behind by the reverse shock (shock $S2$); that is the sound speed inside the plasmon. For this, we identify the impinging flow in the model presented in this Appendix with the original clump. So, if $v_0$ and $\rho_{cl}$ are the launch velocity and density of the clump respectively, then, we take, $v_2= v_0$, $\rho_2=\rho_{cl}$ and $\rho_1$ equal to the density of the ambient medium through which the plasmon is moving $\rho_a$. Then, $c_2$ will be the sound speed inside the plasmon $c$, while $v_c$ (from equation(A8)) will be the initial velocity of the plasmon $v_0$.
Substituting in Eq. (\ref{eqn:c2}) we find,

\begin{equation}
    c=v_0\left( \frac{\gamma-1}{2}\right)^{1/2} \beta.
\label{eqn:cmodel}
\end{equation}


\begin{thebibliography}{}
%
\bibitem[\protect\citeauthoryear{Bally}{2011}]{Bally2011}
{Bally}, J. and {Cunningham}, N.~J. and {Moeckel}, N. and {Burton}, M.~G. and 
	{Smith}, N. and {Frank}, A. and {Nordlund}, A.,
 2011, ApJ,113 , 727

\bibitem[\protect\citeauthoryear{Bally}{2016}]{Bally2016}
{Bally}, J.,2016 ,ARA\&A ,491 , 54

\bibitem[\protect\citeauthoryear{Canto}{1991}]{Canto1991}
{Cant\'o},J.,Raga,A.1991,ApJ., 372,646.

\bibitem[\protect\citeauthoryear{Canto}{1998}]{Canto1998}
{Cant\'o}, J. and {Espresate}, J. and {Raga}, A.~C. and {D'Alessio}, P.
	,1998 , MNRAS,1041 , 296

\bibitem[\protect\citeauthoryear{Castellanos}{2018}]{Castellanos2018}
A. Castellanos-Ram\'\i rez, A. Rodr\'iguez-Gonz\'alez,P. R. Rivera-Ort\'\i z,
A. C. Raga, R. Navarro-Gonz\'alez \& A. Esquivel, 2018 ,RMxAA , 54, 409

\bibitem[\protect\citeauthoryear{Daly}{1994}]{Daly1994}
{Daly}, R.~A. , 1994, ApJ, 38, 426 

\bibitem[\protect\citeauthoryear{Daly}{1995}]{Daly1995}
{Daly}, R.~A. , 1995, ApJ, 580, 454


\bibitem[\protect\citeauthoryear{DeYoung}{1967}]{DeYoung1967}{De Young}, D.~S. and {Axford}, W.~I., 1967, Nat, 129-131, 216

\bibitem[\protect\citeauthoryear{Esquivel et al.}{2010}]{Esquivel2010}Esquivel A., Raga A. C., Cant\'o J., Rodr\'iguez Gonz\'alez, A., L\'opez-C\'amara
D., Vel\'azquez P. F., De Colle F., 2010, ApJ, 725, 1466

\bibitem[\protect\citeauthoryear{Henney}{2009}]{Henney2009}
{Henney}, W.~J. and {Arthur}, S.~J. and {de Colle}, F. and {Mellema}, G.,2009 , MNRAS,157 , 398

\bibitem[\protect\citeauthoryear{Kahn}{1980}]{Kahn1980}
{Kahn}, F.D., 1980, Astron\&Astrophys, 83,303

\bibitem[\protect\citeauthoryear{Kosinski}{2007}]{Kosinski2007}
{Kosi{\'n}ski}, R. and {Hanasz}, M., 2007,MNRAS ,861 , 376

\bibitem[\protect\citeauthoryear{Raga}{1995}]{Raga1995}
Raga,A.,Cabrit, S., Cant\'o,J. 1995,MNRAS, 273,422.

\bibitem[\protect\citeauthoryear{Raga}{1998}]{Raga1998}
{Raga}, A.~C. and {Cant\'o}, J. and {Curiel}, S. and {Taylor}, S.,
 1998, MNRAS , 738, 295

\bibitem[\protect\citeauthoryear{Raga}{2004}]{Raga2004}
{Raga}, A.~C. and {Reipurth}, B., 2004, RMxAA,15 , 40

\bibitem[\protect\citeauthoryear{Taylor}{1992}]{Taylor1992}
{Taylor}, D. and {Dyson}, J.~E. and {Axon}, D.~J.,1992 ,MNRAS ,351 ,255 

\bibitem[\protect\citeauthoryear{Toro et al.}{1994}]{Toro1994} Toro, E. F., Spruce, M., \& Speares, W. 1994, Shock Waves, 4, 25

\bibitem[\protect\citeauthoryear{Ubachukwu}{1991}]{Ubachukwu1991}{Ubachukwu}, A.~A. and {Okoye}, S.~E. and {Onuora}, L.~I., 1991, ApJ, 56, 383

\bibitem[\protect\citeauthoryear{Veilleux}{1993}]{Veilleux1993}
{Veilleux}, S. and {Tully}, R.~B. and {Bland-Hawthorn}, J.,1993 ,AJ , 1318,105 

\bibitem[\protect\citeauthoryear{Yalinewich}{2016}]{Yalinewich2016}
{Yalinewich}, A. and {Sari}, R., 2016, ApJ, 177, 826

\bibitem[\protect\citeauthoryear{Zapata}{2009}]{Zapata2009}
{Zapata}, L.~A. and {Schmid-Burgk}, J. and {Ho}, P.~T.~P. and 
	{Rodr{\'{\i}}guez}, L.~F. and {Menten}, K.~M.,
 2009, ApJ letters, L45, 704

\end{thebibliography}
\end{document}